\documentclass{elsarticle}

\usepackage{hyperref}

\usepackage{amssymb}

\journal{MethodsX}

\usepackage{natbib}
\usepackage[T1]{fontenc}
\usepackage[utf8]{inputenc}

\usepackage{amsmath}

\usepackage{geometry}
\usepackage{fleqn}
\usepackage{graphicx}
\graphicspath{ {./figures/} }
\usepackage{txfonts}

\usepackage{float}
\usepackage{dblfloatfix}
\usepackage{caption}

\usepackage{footmisc}

\usepackage{hyperref}
\usepackage{cleveref}
\Crefname{figure}{Fig.}{Figs.}
\crefname{paragraph}{paragraph}{paragraphs}
\Crefname{paragraph}{Paragraph}{Paragraphs}
\usepackage{xurl}

\usepackage{booktabs,siunitx}

\usepackage{verbatim}

\usepackage{enumitem}
\newlist{longenum}{enumerate}{6}
\setlist[longenum,1]{label=\arabic*)}
\setlist[longenum,2]{label=\alph*)}
\setlist[longenum,3]{label=\roman*)}
\setlist[longenum,4]{label=(\arabic*)}
\setlist[longenum,5]{label=(\alph*)}
\setlist[longenum,6]{label=(\roman*)}

\begin{document}

\begin{frontmatter}
	
	\title{PAD: a graphical and numerical enhancement of structural coding to facilitate thematic analysis of a literature corpus}
	
	\author[label1]{Etienne-Victor~Depasquale\corref{cor1}}
	\cortext[cor1]{Corresponding~author}
	\ead{etienne.depasquale@um.edu.mt}
	\affiliation[label1]{organization={Department of Communications and Computer Engineering, University of Malta},
		city={Msida},
		postcode={MSD2080},
		country={Malta}}
	
	\author[label2]{Humaira~Abdul~Salam}
	\ead{humaira.abdul.salam@desy.de}
	\affiliation[label2]{organization={High-Energy Physics Research (FH) - IT, Deutsches Elektronen-Synchrotron (DESY)},
		city={Hamburg},
		postcode={22607},
		country={Germany}}
	
	\author[label3]{Franco~Davoli}
	\ead{franco.davoli@unige.it}
	\affiliation[label3]{organization={Department of Electrical, Electronic and Telecommunications Engineering and Naval Architecture (DITEN), University of Genoa},
		city={Genoa},
		postcode={16145},
		country={Italy}}
	
	\begin{abstract}
		We suggest an enhancement to structural coding through the use of (a) causally bound codes, (b) basic constructs of graph theory and (c) statistics. As is the norm with structural coding, the codes are collected into categories. The categories are represented by nodes (graph theory). The causality is illustrated through links (graph theory) between the nodes and the entire set of linked nodes is collected into a single \textbf{\textit{directed acyclic graph}}. The number of occurrences of the nodes and the links provide the input required to analyze relative frequency of occurrence, as well as opening a scope for further statistical analysis. While our raw data was a corpus of literature from a specific discipline, this enhancement is accessible to any qualitative analysis that recognizes causality in its structural codes.
	\end{abstract}
	
	\begin{keyword}
		\texttt{Structural coding, thematic analysis, graph theory, nodes, links.}\sep
	\end{keyword}
	
\end{frontmatter}

	
	\section{Motivation}\label{Motivation}
	\subsection{Primary challenge}
	The undertaking of a survey is subject to the risk of degenerating into a ramble around the corpus of literature in scope. The use of data structures, such as tables and lists, gives all stakeholders a sense of value, but these summaries pose a different problem: each such data structure represents a single perspective on the corpus. While these may very well be highly informative and of great value, they fall short in the overarching objective of the survey: a holistic abstraction of the literature that succinctly presents the state of knowledge on the field of study.
	
	This problem may be tackled through the use of analytical methods that are recognized as leading to such an abstraction. We have used \textbf{\textit{thematic analysis}} in surveying, and justify our selection of this form of analysis on:
	\begin{enumerate}
		\item the coherence amongst widely-cited texts \cite{boyatzis_transforming_1998,braun_using_2006,creswell2016qualitative,nowell_thematic_2017,saldana_introduction_2009} in recommending thematic analysis as an introduction to the methods of qualitative analysis, and 
		\item its application in \cite{knowles_exploring_2013}.
	\end{enumerate}
	
	\subsection{Derivative challenge}
	"The excellence of the research [(qualitative analysis)] rests in large part on the excellence of the coding" \cite[{p.~27}]{strauss_qualitative_1987}. Therefore, choice of coding method requires careful consideration of \textbf{\textit{its fitness for the purpose of arriving at the desired abstraction}}. This is the challenge we faced, and this paper is our response to it. Our enhancement of structural coding converts text into:
	\begin{itemize}
		\item \textbf{\textit{numeric data}}, which is then processed using metrics (\cref{MethodStatistics}) that normalize dataset sizes and thereby lead to frequency analysis;
		\item \textbf{\textit{graphical objects}} (\cref{MethodGraphicalMaps}), which are then organized into graphical overviews.
	\end{itemize} .  
	
	These facilitate both objectivity in, and critique of, the insights obtained from the thematic analysis.
	
	\section{Contents}\label{subsec:Contents}
	\begin{enumerate}
		\item In \cref{CodingTheCoreTechnique}, the concept of the "code" is revisited(\cref{subsec:WhatAreCodes}). We declare our use of \textbf{\textit{structural coding}} as the fundamental coding technique and specify the series of questions applied to the raw data(\cref{subsec:HowAreCodesFormed}). This is followed by an explanation of the reasoning that guided the extraction of the codes (\cref{subsec:ApplyingStructuralCode}). 
		\item In \cref{Categorization}, we introduce the \textbf{\textit{node}} and \textbf{\textit{link}} concepts from graph theory (\cref{subsec:OurUseOfNodes}) and proceed to relate the \textbf{\textit{causally-bound codes}} to the node and link.  An algorithm for categorization is presented(\cref{subsec:IterativeProcessOfCategorization}). We then extend the causal linkage inherent to the mined codes to the resulting categories.
		\item \Cref{NodeAttributes} develops the application of nodes in this method by defining attributes that support further processing of the nodes. 
		\item In \cref{MethodGraphicalMaps}, the graphical means for communicating a survey's results are described.
		\item Statistics produced by the method, as well as intuition on their significance, are described in \cref{MethodStatistics}.
		\item An explanation of how the method facilitates thematic analysis is presented in \cref{Deriving themes}.
		\item Observations on the application of the method are presented through case notes in \cref{CaseNote}, where we draw an outline of how this method was applied to a recent work.
		\item We conclude by identifying benefits and limitations (\cref{subsec:BenefitsAndLimitations}) of the method and summarize the prescriptions of the method in \cref{subsec:MethodSummary}.
		
	\end{enumerate} 
	
	\section{Coding: the core technique} \label{CodingTheCoreTechnique}
	\subsection{What are codes?}\label{subsec:WhatAreCodes}
	\textbf{Codes are terse, dense, textual representations of a verbose articulation of a concept embedded in the raw data}.  The prescriptions of sound qualitative analysis for systematic review require a choice of \textbf{\textit{coding method}}. The process of coding collates the diversity of the surveyed set of papers through the formation of smaller \textbf{\textit{codes}}. We refer to each paper included within the scope of the survey as a \textbf{\textit{research unit}} (RU), i.e. a publication (\textbf{\textit{excluding}} surveys) in conference proceedings and journals.
	Codes must have the following characteristics.
	\begin{enumerate}
		\item They must be semantically rigorous i.e. the meaning a code represents must be clear and use (application) of the code must be unconfusable.
		\item They must be universally applicable across RUs, i.e. they must provide a uniform means of dissecting publications. Use of more than one coding system (i.e. two or more non-universal coding systems) may create a split in the coded data with incomparable parts across the split.
	\end{enumerate}
	\subsection{How are codes formed?}\label{subsec:HowAreCodesFormed}
	We satisfy these two requirements through the elemental coding method of \textbf{\textit{structural coding}}. Structural coding poses \textbf{\textit{a series of questions relevant to the inquiry in hand}} and is well suited to any problem which can be described using a standardized set of questions. We deconstruct the problem of literature review into a standardized question-and-answer protocol that can be directly converted into a structural coding approach. The questions are the following:
	\begin{enumerate}
		\item What is the problem which the researcher(s) saw as an opportunity for study?
		\item What approach(es) did the researchers take in an attempt to solve the problem?
		Or: \textit{how did they go about it ... what did they \textbf{do}}?
		\item What development(s) and/or contributions derive from the researcher(s) work?
	\end{enumerate}
	
	The corresponding structural codes are:
	\begin{enumerate}
		\item \textbf{P}roblems (or challenges)
		\item \textbf{A}pproaches
		\item \textbf{D}evelopments (or contributions)
	\end{enumerate}
	We refer to this protocol as the \textbf{\textit{PAD review protocol}}. Each RU (paper) is mined for:
	\begin{enumerate}
		\item “the problem which the researcher(s) saw as an opportunity for study” (P-codes),
		\item the “approach(es) … [taken by] the researchers … in an attempt to solve the problem” (A-codes) and
		\item the “developments(s) and/or contributions deriv[ing] from the researcher(s) work” (D-codes).
	\end{enumerate}
	
	(quoted text is taken from the questions enumerated above).

	\subsection{Applying the structural code: identifying problems, approaches and developments} \label{subsec:ApplyingStructuralCode}
	We now express a generalized understanding of the reasoning we follow to identify the problems, approaches and developments. In terms of thematic analysis, this is the role which an expert would play to transform the raw data into codes.This applies a first abstraction to the RUs:
	\begin{itemize}
		\item from detail specific to the RU
		\item to abstractive problem-, approach- and development-codes
	\end{itemize}
	
	that meaningfully represent the individual RUs. 
	\subsubsection{Problem codes}
	Research is rooted in “the problem which the researcher(s) saw as an opportunity for study”. Key techniques in identification of the problem include:
	\begin{itemize}
		\item a focus on the abstract and the introduction,
		\item making a high-level summary, e.g. by answering questions such as: what is this paper {trying to solve}/{concerned with}, and
		\item recognition of revealing phrases like "in this paper" or "in this work".
	\end{itemize}
	When the scope of a survey is relatively narrow, the problems in the set may not be fully independent of one another. They may diverge from one another only as \textbf{\textit{aspects}} (we could also say that they are \textbf{\textit{derivatives}}) \textbf{\textit{of a core challenge}}. In such a case, each RU would be rooted in this core challenge, but the derivative problems (the P-codes) addressed differ from one RU to another.
	\subsubsection{Approach codes}
	\textbf{\textit{Approach}} is used in the sense defined by the Oxford English Dictionary as “\textit{figurative. A way of considering or handling something, esp. a problem}.” This meaning is broad yet it is purposely cited here to convey some of the depth of the difficulty we faced in assessing whether an observation candidate as an approach should be included (or not). Key techniques in identification of the approach include:
	\begin{itemize}
		\item a focus on the method section
		\item \textbf{\textit{choices}} that focus technique
	\end{itemize} 
	Codes of this type are the hardest to extract from RUs, as they are heavily dependent on a good grasp of the research space. This difficulty is severe enough to count as a limitation of our method (see \cref{subsec:BenefitsAndLimitations}). We emphasize that the “way of considering or handling something” is never (we dare say) obtained through a single approach, but, rather, through an entire delta (fan-out) of component approaches that are combined to bring efforts to a yield: the development.
	\subsubsection{Development codes}
	\textbf{\textit{Developments}} include what are commonly referred to as contributions, but we can afford to broaden our scope for inclusion of less significant products of research. Therefore, we go beyond simply parsing RUs in search of the familiar “contributions of our paper” or “in this paper” phrases, and glean those useful bits that, taken alone, do not qualify as the scope of a paper.
	
	\section{Categorization: clustering the codes} \label{Categorization}
	The act of coding is carried out as an intermediate step on the way towards \textbf{\textit{categorization}}. We first introduce our use of the concept of the \textbf{\textit{node}}. We then proceed to a detailed treatment of how, through iterations, a set of problem-, approach- and development-codes can be categorized.
	
	\subsection{Nodes - our use of a graph theory concept to complement structural coding}\label{subsec:OurUseOfNodes}
	We use the concept of \textbf{\textit{node}} from graph theory. The product of mining an RU is one or more P- , A- and D-codes. Each code is encapsulated within a node. Since problem, approach and development are linked to one another in \textbf{\textit{a causal chain}}, \textbf{\textit{this causal relationship can be represented through linked nodes.}} Each such causal chain is represented by three nodes; thus we call it a \textbf{\textit{triad}}. Our application of the node concept is beneficial for the following reasons:
	
	\begin{enumerate}
		\item An RU's collection of triads (of nodes) is a synthetic representation of the RU, that facilitates a good apprehension thereof.
		\item These characterizing triads facilitate the task of locating this RU within the greater landscape of research.
	\end{enumerate}  
	
	\subsection{The iterative process of categorization} \label{subsec:IterativeProcessOfCategorization}
	For every RU in the corpus, we iterate through a number of steps to categorize a code.
	\begin{longenum}
		\item Given an RU, a set of \textbf{\textit{problem, approach and development}} codes is collected and added to a pool of ungrouped codes.
		\item For every code, associate the code, say $\alpha$, with a numerical, integral co-attribute, say $a$, and add both the text and number as attributes of a node. The code is, therefore, an aspect, or \textbf{\textit{attribute}} of the node.
		\item Seek a pre-extant code (of the same type, i.e. P/A/D) nearest in meaning, say code $\beta$, with numerical, integral co-attribute $b$.
		\item If there is no neighbor (in meaning) sufficiently close to group $\alpha$ with, 
		\begin{longenum}
			\item then this iteration of categorization terminates for code $\alpha$ and it is kept within the pool of ungrouped codes.
			\item else:
			\begin{longenum}
				\item If $\beta$ is \textbf{\textit{not}} already grouped within a category,
				\begin{longenum}
					\item then:
					\begin{longenum}
						\item create a category, with proximating meaning $\Gamma$ common to both $\alpha$ and $\beta$, and with numerical, integral co-attribute $C$ (upper case, to distinguish this as a category);
						\item replace numerical co-attribute $b$ by $C.1$ while $a$ is replaced by $C.2$, and
						\item add both the text $\Gamma$ and number $C$ as attributes of a \textbf{\textit{category-node}}.
					\end{longenum}
					\item else: (i.e. if $\beta$ \textbf{\textit{is}} already grouped within a category)
					\begin{longenum}
						\item revise the scope (therefore, the text) of the proximating meaning to take $\alpha$ into account;
						\item if the proximating meaning (i.e. $\Gamma$) cannot be reasonably revised to include code $\alpha$,
						\begin{longenum}
							\item either spawn a new category, containing codes $\alpha$ and $\beta$;
							\item or keep $\alpha$ as an orphan (i.e. without a grouping category), within a pool of ungrouped codes.
						\end{longenum}
					\end{longenum} 
				\end{longenum}
			\end{longenum}
		\end{longenum} 
	\end{longenum}
	
	Note that:
	\begin{itemize}
		\item we refer to the proximating meaning as the \textbf{\textit{category code}};
		\item if $\beta$ had already been grouped within a category, its numerical co-attribute would already be of the float type (i.e. include a decimal separator), as it would have been previously transformed from numerical integer $b$ to some float type that includes the digit(s) of its grouping category. 
	\end{itemize}
	
	The category code describes the salient meaning common to all codes grouped under it. It is the abstraction that sacrifices some detail for the sake of facilitating a holistic view of the research space. The resulting categorization is conducive to (a) assimilation by a viewer, as well as (b) further rationalization.
	
	\section{Node attributes}\label{NodeAttributes}
	The process of categorization results in a set of problem-category-, approach-category- and development-category-nodes. Each type of category-node groups nodes of the same type (P/A/D). This section describes the attributes of the individual nodes and those of the category-nodes. 
	
	\subsection{Node attributes}
	\begin{enumerate}
		\item a short textual description – the \textbf{\textit{code}} – of the observed problem/approach/development and
		\item a unique alphanumeric identifier (a \textbf{\textit{label}}) of the form P/A/Dx.yz. In this labelling scheme,
		\begin{enumerate}
			\item P, A and D represent the structural codes,
			\item x is a single digit that represents the observed problem/approach/development (P/A/D) category, 
			\item y is a single digit that represents an observed sub-category of the P/A/D category, if any such sub-category is observed, and
			\item z is a single digit that represents the individual, observed problem/approach/development.
		\end{enumerate}
		We chose the notation x.yz instead of x.y.z to type the variable represented by this identifier as a numeric float type.
	\end{enumerate}
	
	\subsection{Category-node attributes}
	\begin{enumerate}
		\item a short textual description – the \textbf{\textit{category code}} – of the observed P/A/D category;
		\item a unique alphanumeric identifier (a \textbf{\textit{label}}) of the form P/A/D$x$. In this labelling scheme,
		\begin{enumerate}
			\item P, A and D represent the structural codes,
			\item $x$ is an integer that represents the observed problem/approach/development category.
		\end{enumerate}
		\item a set of member nodes.
	\end{enumerate}
	Where major sub-clusters need to be identified, we extend the labelling to P/A/Dx.y ,where y is a second integer representing the sub-cluster. In these cases, as we show in the next sub-sub-section, the node identifier takes the form P/A/Dx.yz.
	
	\section{Graphical maps of the research space} \label{MethodGraphicalMaps}
	The workings of the method culminate in the production of several graphic devices, which we describe below.
	\subsection{The causality DAG}
	The P-, A- and D- category-nodes are linked according to the triads mined from the RUs, to produce a graphic device which may be tersely and aptly referred to as a \textbf{\textit{directed acyclic graph (DAG) of causality}} - the causality DAG. This graphic device is an important part of the product of our method. It shows a bird’s-eye view of the dynamics of research in our chosen scope. 
	
	In particular, the DAG is an encoding of surveyed research units, that strives to relieve a profile not only of current knowledge (the \textbf{\textit{developments}}) but also of what has been found a fruitful pursuit (the \textbf{\textit{approaches}}) thereof. This is obtained through the relationships that are illustrated in the DAG between the problems addressed, the approaches to solutions, and the knowledge obtained in pursuit of solutions.
	
	The DAG may show links between \textbf{\textit{category-nodes}}, rather than individual nodes, to minimize clutter and improve readability. Such a reduction is strongly dependent on truly representative categorization (see \cref{Categorization}), through groupings meaningful to the survey's scope. In this form of the DAG, the inter-category links are \textbf{\textit{aggregators}}: they include all links between any two nodes within their respective categories. \textbf{\textit{Line thickness}} is an excellent way to represent the size of the set of aggregated links.
	
	\subsection{Triads graphic}
	We also highlight triads. Triads are represented by lines passing through a single combination of a single problem, a single component of approach and a single development to which the approach (component) led (not necessarily on its own; indeed, rarely so). The triads graphic identifies the dynamics of research at a glance, indicating the most highly used triads by line thickness. 
	
	As with the DAG (and for the same reason), the triads graphic may show causal bindings between category-nodes, not individual nodes. Each triad is a serial connection of one link from a P-category to an A-category, and one link from the latter A-category to a D-category. Each triad is illustrated with a bend where links meet on A-category nodes.
	
	\subsection{P-A dyads graphics: challenges (P) and associated approaches (A)}
	The causality DAG's section showing P-A category links is partitioned into a set of P-A dyad-graphics. Each such graphic illustrates the set of all approach components that have been used to tackle a particular problem category. The P-A dyads graphics complement the causality DAG and the triads graphic by a focus on how individual challenges have been tackled in published research. Each P-A dyads graphic provides, from the perspective of a specific challenge, the same information on frequency (of applied approaches) as the (global) causality DAG.
	
	\subsection{Taxonomies of problems, approaches and developments}
	Within each of the three divisions (i.e. problems, approaches and developments) of categories, frequency of occurrence of categories is expected to communicate meaningful information about the state of the art. Furthermore, within each category-division, it may be possible to find relationships between categories that convey additional meaning and encourage structural formations that gather the categories into \textbf{\textit{taxonomies}}. For example: if two categories of approaches are proximal in meaning, a super-category might be formed that abstracts the differences in meaning between the two and represents them both from the perspective of the common, salient meaning.
	
	\section{Statistics}\label{MethodStatistics}
	The frequency of occurrence of aspects of data collected is examined here. We combine the structural codes in various ways in order to obtain useful statistics for a quantitative grasp of the field. In the following sub-sections, we suggest and describe how these statistics can be collected from the data.
	
	\subsection{Frequency of occurrence of a challenge category in RUs}\label{subsubsec:FreqOccurrenceChallenges}
	We start by proposing the following simple relationship.
	\begin{equation}
		F_{P_k} = \frac{\sum_{j=1}^{N_{RU}}P_k^{(j)}}{{N_{RU}}}
	\end{equation}
	
	where:
	\begin{enumerate}
		\item $P_k^{(j)}$ is a binary variable that represents the presence (or lack thereof) of a problem in category $P_k$,
		\item within a single RU $RU_j$ over the corpus of $N_{RU}$  unique RUs.
	\end{enumerate}
	
	This simply indicates the number of times in which a (derivative) challenge-category appears within the RUs in the corpus, as a fraction of the total number of RUs. The numerator of $F_{P_k}$ is incremented once by $RU_j$ for a given $P_k$ if a problem in this category is tackled in $RU_j$.
	
	\subsection{Research interest: frequency of occurrence of a challenge category among all occurrences of challenge categories}\label{MethodResearchInterest}
	Here, we define \textbf{\textit{research interest}}, denoted by $R_{P_k}$, in a given challenge (problem) category $P_k$, as its frequency of occurrence within the set of all the challenges tackled in all research units. It is computed as the total number of times in which problems in category $P_k$ have been tackled in RUs, as a fraction of the sum of the total number of times in which (problems in) all observed challenge categories have been tackled. We suggest this as a metric of the attention, or research interest, which this challenge is receiving. The numerator of $R_{P_k}$ is incremented once for $RU_j$ for a given $P_k$ if a problem in this category is tackled in $RU_j$.
	\begin{equation}
		R_{P_k}= \frac{\sum_{j=1}^{N_{RU}}P_k^{(j)}}{\sum_{i=1}^{N_{P}}\sum_{j=1}^{N_{RU}}P_i^{(j)}}
	\end{equation}
	
	where:
	\begin{enumerate}
		\item $P_k^{(j)}$ is a binary variable that represents the presence (or lack thereof) of a problem in category $P_k$,
		\item within a single RU $RU_j$ over the corpus of $N_{RU}$  unique RUs,
		\item with a total of $N_P$  unique, identified challenge/problem categories.
	\end{enumerate} 
	
	Note that:
	\begin{equation}
		\frac{F_{P_k}}{R_{P_k}} = \frac{\sum_{i=1}^{N_{P}}\sum_{j=1}^{N_{RU}}P_i^{(j)}}{N_{RU}}
	\end{equation}
	This ratio is a constant; therefore, both $F_{P_k}$ and $R_{P_k}$ have identical distributions, but the statistics differ. Specifically, the ratio $\frac{F_{P_k}}{R_{P_k}}$ is the average number of challenges tackled per RU.
	
	\subsection{Metric of observed approach diversity: Weighted challenges}\label{MethodWeightedChallenges}
	We measure the diversity of approaches through which a challenge is tackled.  The set of all unique problem-approach (PA) pairs (dyads) in the triads is collected first. Therefore, a particular PA dyad is counted once, regardless of the number of occurrences of that dyad. Then, for each problem, we add up the total number of dyads within which that problem is found. We suggest a normalized diversity metric, $W_{P_k}$, as follows:
	\begin{equation}
		W_{P_k}=\frac{\sum_{j=1}^{N_{PA}}P_k^{(j)}}{\sum_{i=1}^{N_{P}}\sum_{j=1}^{N_{PA}}P_i^{(j)}}
	\end{equation}
	
	where:
	\begin{enumerate}
		\item $P_k^{(j)}$  is a binary variable that represents the presence (or absence) of a problem category $P_k$,
		\item within a single P – A dyad $PA_j$, over the set of all $N_{PA}$ unique P – A dyads within the corpus,
		\item with a total of $N_P$  unique, identified challenge/problem categories.
	\end{enumerate}
	
	\subsection{Frequency of occurrence of an approach category among all occurrences of approach categories}\label{FrequencyACatAmongAllACats}
	Unlike problems in challenge categories, approaches within an approach category do not, most commonly, mutually exclude one another. Therefore, the count of occurrences of an approach category may be incremented more than once per RU and thus a metric like $F_{P_k}$ lacks a good normalization basis. As it is useful to learn how widely exploited an approach is, a different normalization basis must be selected. We therefore use a metric somewhat similar to $R_{P_k}$, i.e. frequency of occurrence of an approach category in all RUs, among the set of all occurrences of approach categories in all RUs. We obtain the metric $R_{A_k}$, as follows:
	
	\begin{equation}
		R_{A_k}=\frac{\sum_{j=1}^{N_{RU}}\sum_{m=1}^{\lvert A_k \rvert}A_{k_m}^{(j)}}{\sum_{i=1}^{N_A}\sum_{j=1}^{N_{RU}}\sum_{m=1}^{\lvert A_i \rvert}A_{i_m}^{(j)}}
	\end{equation}
	
	where:
	\begin{enumerate}
		\item $A_{k_m}^{(j)}$ is a binary variable that represents the presence (or lack thereof) of \textbf{\textit{approach}} $A_{k_m}$,
		\item where  $A_{k_m}$ is a member of \textbf{\textit{approach category}} $A_k$, of cardinality $\lvert A_k \rvert$,
		\item within a single RU $RU_j$,
		\item over the corpus of $N_{RU}$ unique RUs,
		\item with a total of $N_A$  unique, identified approach categories.
	\end{enumerate}
	
	Note that occurrence of category $A_k$ within an RU is counted as many times as its members $A_{k_m}$ appear in the RU.
	
	\subsection{Metric of utility of an approach: Weighted approaches}\label{MethodMetricUtilityApproach}
	We also analyze approaches in terms of their utility, i.e. how useful they are in the overall motion between problems and developments, and denote this metric as $U_{A_k}$. The numerator is incremented each time a particular approach is a component of a triad within an RU. Therefore, a single RU may increment the metric several times. The utility metric of a specific approach \textbf{\textit{$A_k$}} is the normalized metric:
	
	\begin{equation}
		U_{A_k}=\frac{\sum_{j=1}^{N_{RU}}\sum_{l=1}^{N_{triads_{RU_j}}}A_k^{(l)}}{\sum_{i=1}^{N_{A}}\sum_{j=1}^{N_{RU}}\sum_{l=1}^{N_{triads_{RU_j}}}A_i^{(l)}}
	\end{equation}
	
	where:
	\begin{enumerate}
		\item $A_k^{(l)}$ is a binary variable that represents the presence (or lack thereof) of approach category $A_k$,
		\item within any of the $N_{triads_{RU_i}}$ triads in a single research unit $RU_j$,
		\item over the corpus of $N_{RU}$ unique RUs,
		\item with a total of $N_A$  unique, identified approach categories.
	\end{enumerate}
	
	We emphasize that a single research unit may be described by several such triads that include approach $A_k$.

	\subsection{Frequency of occurrence of categories of development}
	Development statistics are distributed thinly unless developments are categorized. However, when grouped into meaningful clusters (categories), conclusions can be drawn about the frequency with which developments take place in sub-spaces of this research space. We obtain $R_{D_k}$, the normalized frequency of occurrence of categories, as follows:
	
	\begin{equation}
		R_{D_k}=\frac{\sum_{j=1}^{N_{RU}}\sum_{m=1}^{\lvert D_k \rvert}D_{k_m}^{(j)}}{\sum_{i=1}^{N_D}\sum_{j=1}^{N_{RU}}\sum_{m=1}^{\lvert D_i \rvert}D_{i_m}^{(j)}}
	\end{equation}
	
	where:
	\begin{enumerate}
		\item $D_{k_m}^{(j)}$ is a binary variable that represents the presence (or lack thereof) of \textbf{\textit{development}} $D_{k_m}$,
		\item where  $D_{k_m}$ is a member of \textbf{\textit{development category}} $D_k$, of cardinality $\lvert D_k \rvert$,
		\item within a single RU $RU_j$,
		\item over the corpus of $N_{RU}$ unique RUs,
		\item with a total of $N_D$  unique, identified development categories.
	\end{enumerate}

	Thereby, a prospective researcher is guided through grounded insight into works covering this space.
	
	\section{Deriving themes} \label{Deriving themes}
	The concept of a theme is consistently described as a \textbf{\textit{pattern}} that emerges from the raw data; see, e.g. \cite[{p.~82}]{braun_using_2006} and \cite[{p.~4}]{boyatzis_transforming_1998}. The PAD enhancement of structural coding facilitates thematic analysis by accentuating themes through the abstractive function of causally-bound categories of codes. Both quantitative and qualitative aspects of analysis are possible.
	\subsection{Quantitative analysis}
	The \textbf{\textit{results}} of application of the PAD method are powerfully conducive to a quantitative aspect of thematic analysis: 
	\begin{enumerate}
		\item the frequency of individual categories (i.e. whether problem-, approach- or development-categories) is itself meaningful; 
		\item it is possible to attempt an interpretation of the frequency of occurrence of pairs (\textbf{\textit{dyads}}) of problem – approach categories, and  
		\item the frequency of a triad within the overall set identified, is highly representative of the state of the art. We find it more summative to observe triads of category-nodes than triads of individual nodes, as the latter disperse frequency of occurrence too thinly. Categories act as bins that aggregate frequency usefully.
	\end{enumerate}
	
	\subsection{Qualitative analysis}
	Qualitative analysis inheres in the very \textbf{\textit{processes}} of coding and categorization, and it is further accentuated by the causal linkages between the categories of the structural code. The observed categories, dyads, triads and their relative frequencies are fertile grounds for \textbf{\textit{grounded reflection}} about the state of research. Some examples are given in the context of the case notes (\cref{subsec:EmergenceOfThemes}).
	
	This culminates our thematic analysis.
	
	\section{Case notes} \label{CaseNote}
	We have used the PAD method during a survey of research into power modeling and measurement in virtualized environments. The surveyed body of papers was gathered from the ACM, IEEE and other sources (the “corpus”). The corpus is our raw qualitative data.
	
	\subsection{Paper selection criteria}
	Every surveyor will decide on the relevance (in/out of scope) of a paper by considering certain criteria. A generally valid  criterion is to exclude other surveys from the corpus, as a survey is not itself comparable with the works within its scope. This does not exclude surveys from consideration, since a prospective surveyor would be well advised to learn about what ground other surveyors have covered and results obtained from their coverage. However, the surveys do not themselves constitute raw qualitative data: they contain results obtained from the processing of raw data.
	
	In our case, the criterion that proved most effective in sorting RUs into relevant or irrelevant was the challenge undertaken (or what we may now call: \textbf{\textit{the P-node}}). It seems fair to extrapolate this observation to surveys in general, or at the very least, to a major category of surveys. Many surveys of the state of the art in some field of a discipline are oriented towards the progress achieved in tackling the field's challenges. Hence, the generalization we suggest would hold for this category of surveys. An illustration of how the field of study emanates categories of challenges (the P-category-nodes) is  \href{https://github.com/ijqm/pad/blob/main/The Core Challenge and its Derivatives.pdf}{included among the files on our study's GitHub site}\footnote{\url{https://github.com/ijqm/pad/blob/main/The Core Challenge and its Derivatives.pdf}}
	
	\subsection{Identifying codes and categorizing them}
	Problem codes are frequently identified using the abstract and certain key phrases like "in this paper", or "this paper", or "in this work". As the number of RUs perused increases, nuances start emerging and codes that at first appeared to be somewhat distinct are recognized as factually indistinguishable, even before actual categorization begins. These codes are assigned a number and integrated with their numbers in nodes (node = code + number).
	
	As with problem codes, the abstract of an RU is a good source of development codes. At least some development codes are usually self-evident (i.e. \textbf{\textit{semantic}}) here, since researchers are keen to point out their primary contribution(s) (developments) and hold on to readers' notoriously volatile attention. However, thorough harvesting was only obtained with an organic growth in familiarity with the field, as contributions that were less conspicuous or weighty started to emerge as more RUs were perused. These secondary contributions were distributed throughout the length of papers. Therefore, the pace of harvesting development codes was slower than its problem-codes counterpart, as their transparency, number and distribution were less favourable.
	
	Most problematic were the approach codes. These require a difficult movement from the general ("a way of considering or handling something, esp. a problem") to the domain-specific. A useful generalization is that these codes describe the empirical setup and RU sections titled "method" or "methodology" are good sources of codes. Indeed, this would be the "way of ... \textbf{\textit{handling}} something." However, there is the rather \textbf{\textit{latent}} aspect of researchers' "way of \textbf{\textit{considering}} ... a problem". For example, we observed that models developed are at least in part the result of an approach towards modelling, and justification of the selected form of power model is strongly dependent upon the inputs and parameters of operation.
	
	Among the files on our study's GitHub site:
	\begin{itemize}
		\item we cross-reference a comprehensive list of \href{https://github.com/ijqm/pad/blob/main/NodeAttributes.xlsx}{harvested codes(labels) and node numbers}\footnote{\url{https://github.com/ijqm/pad/blob/main/NodeAttributes.xlsx}}, and
		\item we present \href{https://github.com/ijqm/pad/blob/main/Triads.xlsx}{all node triads}\footnote{\url{https://github.com/ijqm/pad/blob/main/Triads.xlsx}} harvested from the corpus.
	\end{itemize}
	
	Categorization of codes proceeded as described in \cref{Categorization}, with P-, A-, and D-nodes handled separately. While the final categorization of each code type required several iterations over the full set of codes, it was a comparatively lightly taxing endeavour. The process of coding directly led our coding work through the considerations that drive categorization: inspection, comparison, contrast, etc., all while noting down codes. 
	
	\begin{figure*}[!b]
		\centering
		\captionsetup{justification=centering}
		\includegraphics[width=1.0\textwidth]{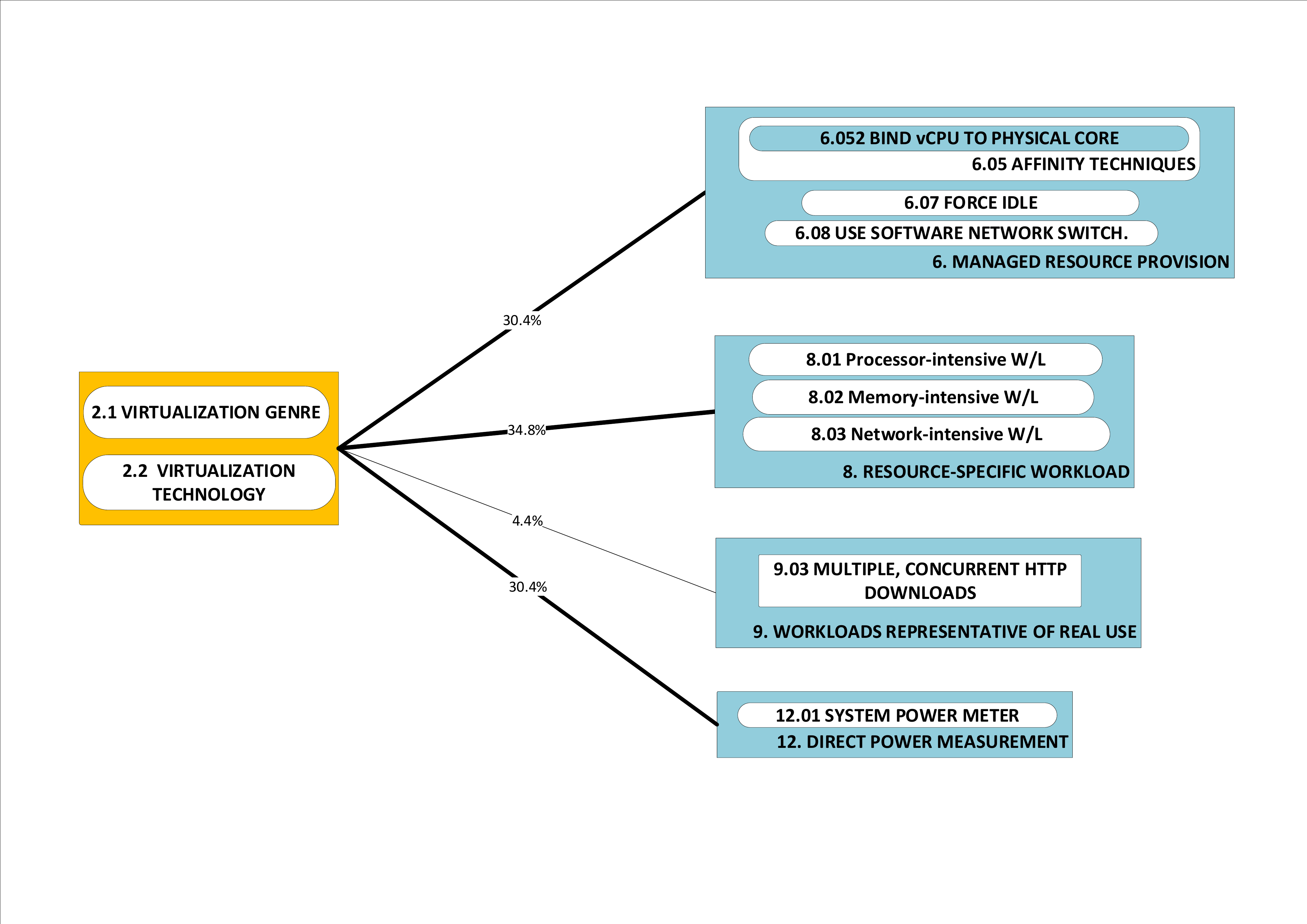}
		\caption{Frequency of occurrence of approaches to tackling P1, shown in line thickness and as percentage}
		\label{fig:p1Dyads}
	\end{figure*}
	
	\begin{figure*}[!t]
		\centering
		\captionsetup{justification=centering}
		\includegraphics[width=1.0\textwidth]{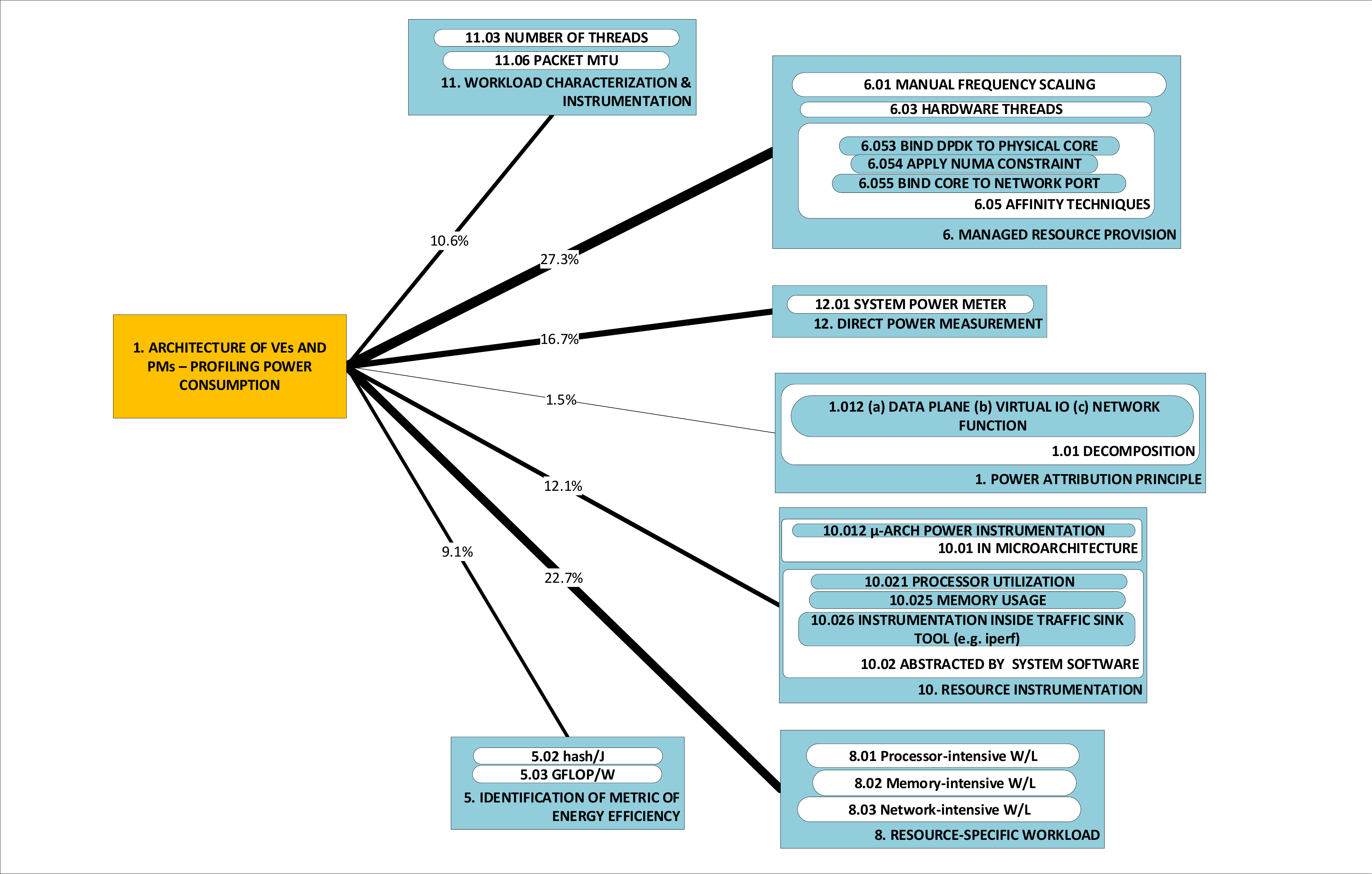}
		\caption{Frequency of occurrence of approaches to tackling P2, shown in line thickness and as percentage}
		\label{fig:p2Dyads}
	\end{figure*}
	
	\subsection{Graphical maps}
	The causality DAG derived from our study is \href{https://github.com/ijqm/pad/blob/main/CausalityDAGWithLinksBetweenCNodesOnly.pdf}{included among the files on our study's GitHub site}\footnote{\url{https://github.com/ijqm/pad/blob/main/CausalityDAGWithLinksBetweenCNodesOnly.pdf}}. The concern with problems in category P7 (resource use and measurement) stands out in the DAG. The primary category of approaches taken to tackle this problem is A10 (through instrumentation of computing resources).
	
	The triads graphic derived from the study is \href{https://github.com/ijqm/pad/blob/main/TriadsWithLinksBetweenCNodesOnly.pdf}{also included}\footnote{\url{https://github.com/ijqm/pad/blob/main/TriadsWithLinksBetweenCNodesOnly.pdf}} on our study's GitHub site . In this graphic, it can be seen that the development that has emerged most frequently out of the P7-A10 dyad, is that of models that regress use of computing resources onto a linear relationship with power consumption. 
	
	Finally, we include herein two examples of dyad graphics from our study. We have chosen direct inclusion of these graphics as they are the least dense of all the devices and thus well suited to serving as a handy example. \Cref{fig:p1Dyads} shows that the challenge of obtaining broad indications of the effect of architecture on power consumption, is tackled through a wide variety of approaches. \Cref{fig:p2Dyads} shows that cross-comparison of power consumption of virtualization genres and technologies (the challenge) has been primarily tackled using workloads that target specific resources (34.8\%), rather than workloads that represent real use (4.4\%).
	
	Several other graphic devices were compiled using this method's products. It has proven to be well suited to the compilation of taxonomies of challenges, approaches and developments in the surveyed field. 
	
	\subsection{The emergence of themes}\label{subsec:EmergenceOfThemes}
\begin{figure*}[!t]
		\centering
		\captionsetup{justification=centering}
		\includegraphics[width=\textwidth]{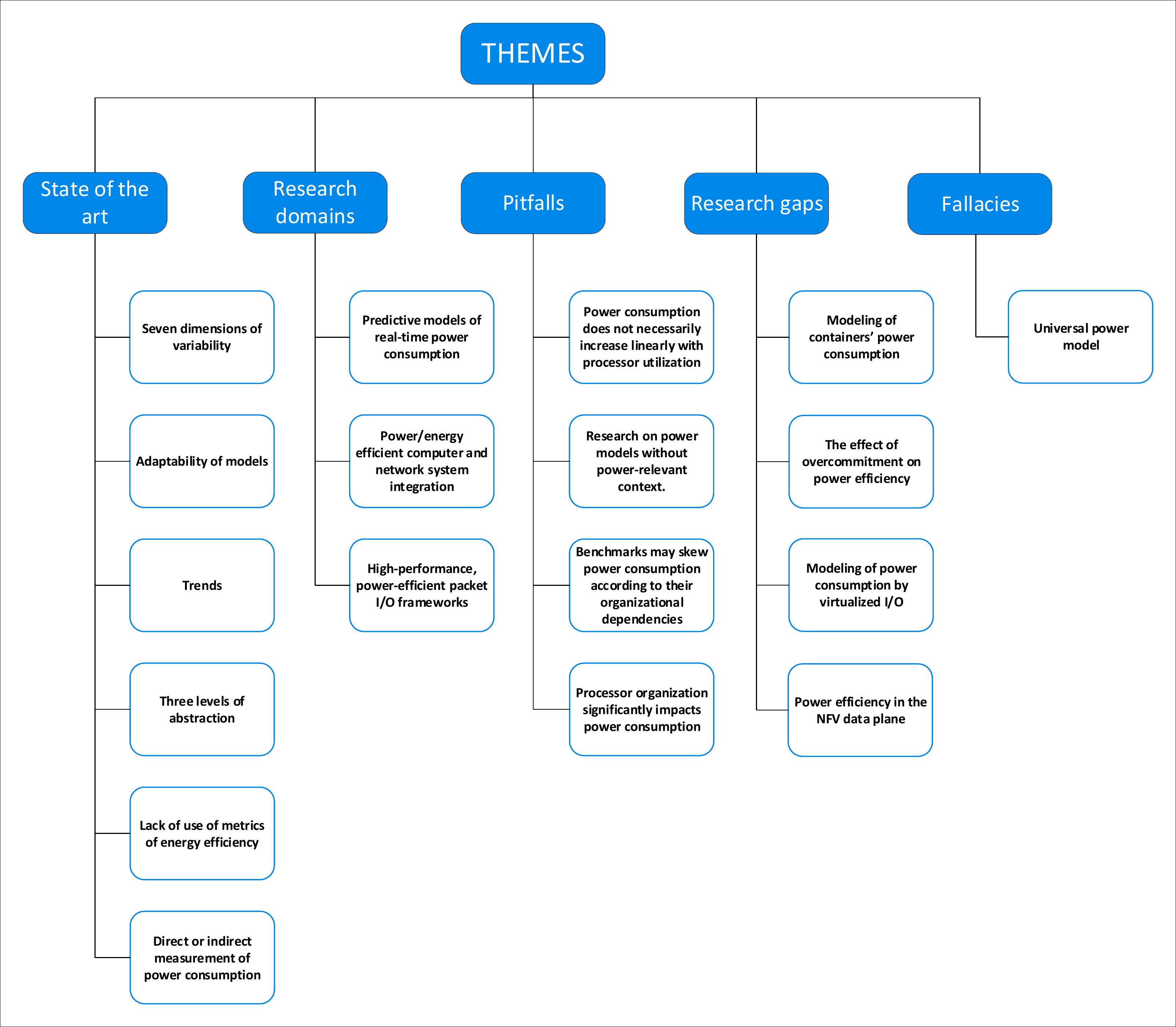}
		\caption{An overview of the qualitative aspect of thematic analysis of the research space}
		\label{fig:themesTree}
	\end{figure*}

	Our knowledge of the surveyed domain developed as we progressed in parsing RUs and discussing the coding and categorization thereof. These processes - coding, categorization, linking and discussion - led to a thorough qualitative analysis, which was itself amenable to classification. Thus, we were able to give a bird's-eye view (see \Cref{fig:themesTree}) of:
	\begin{itemize}
		\item the state-of-the-art, including trends in research;
		\item research gaps;
		\item pitfalls and fallacies, and
		\item domains which demand research into models and measurement of power consumption.
	\end{itemize} 
	
	\section{Conclusion}
	\subsection{Benefits and Limitations} \label{subsec:BenefitsAndLimitations}
	Thematic analysis with structural coding provides at least four distinct benefits.
	\begin{enumerate}
		\item It facilitates \textbf{\textit{the discovery of patterns}} in research, through the causality DAG. The causality DAG is a map that illustrates a structural encoding of primary research sources. This is a robust base for the development of a set of themes that represents the ongoing effort of discovery within the field of study. Patterns may be found in the horizontal dimensions of the DAG. A pattern in the horizontal dimension can consist of a PAD triad that recurs several times in the data. It can also consist of a group of triads that share both the P-category-node and the A-category-node. Such a group of triads comprises a variety of developments that arise out of the same approach to the same problem. A looser but nonetheless interesting grouping comprises those triads that share the P-category-node. This would be useful to a researcher seeking to learn how others who have addressed the same problem.
		\item It facilitates \textbf{\textit{the evaluation of the novelty of research proposals}}. Whether the proposal has reached the stage of problem identification or selection of approach, the DAG is a useful tool in the assessment of the likelihood of developing successful research out of the proposal. This follows because inspection of the map leads to an indication of the density of research in and near the space under consideration for research.
		\item Unlike simpler, ad hoc surveying, the processes used by this method are open to scrutiny through identifiable and tangible proceedings. Whilst still a subjective (see limitations) method, its workings are more amenable to a reader’s analysis than other, less open techniques.
		\item Structural coding leads the surveyor to process his/her corpus systematically. The generalizable, parsing technique reduces the verbosity of text to a regular set of codes that aptly and succinctly describe the unit of research.
	\end{enumerate}
	
	Two limitations have emerged.
	\begin{enumerate}
		\item \textbf{\textit{Clustering coarsens resolution.}} The causality DAG links category-code-nodes. This results in an apparent linkage between any node within one category-node and any node within the category-node at the other end of the link. However, this is not necessarily reflected in the RUs we have mined and is a result of the loss of resolution that accompanies clustering. The effects of this generalization can be mitigated by ensuring that the clustering notion has narrowly defined meaning. Such narrowness in meaning reduces the scope for interpretative error.
		\item \label{iterations} \textbf{\textit{The need for multiple iterations.}} Identification of codes is limited by the reviewer’s breadth of vision of the field under review. Since the work may be carried out with the very purpose of gaining a broad view, it may seem that a Catch 22 is embedded in the method. This is not expected to be as frustrating as it may appear. It is expected that a reviewer has some background within the field. The same ability used to search the field may be exploited during an initial coding \textbf{\textit{iteration}}. As the review proceeds, the reviewer’s breadth of vision expands and the set of codes is grown through refinement of existing codes and addition of new ones. The limitation may be experienced in any of the P-, A- and D- categories. In particular, during this survey, perception of approaches improved as more RUs were parsed. Indeed, this progression is recognized as part of the labor of research: it is partly undertaken during an initial familiarization \cite[{p.~87}]{braun_using_2006} and partly as a cyclical re-evaluation of RUs in the light shed by discovery of new codes. \cite[{p.~8}]{saldana_introduction_2009}.
	\end{enumerate}
	
	\begin{figure*}[!t]
		\centering
		\captionsetup{justification=centering}
		\includegraphics[width=\textwidth]{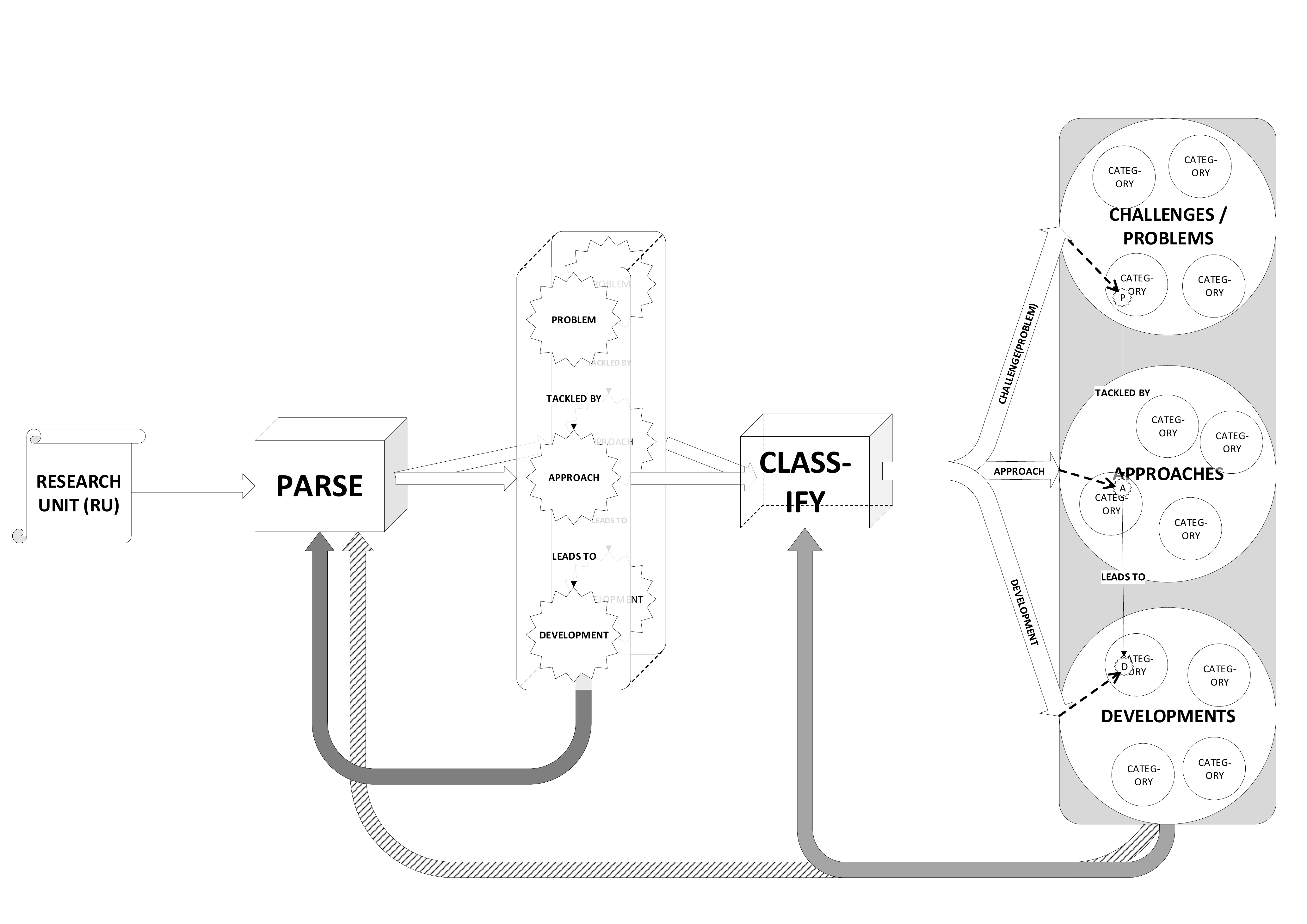}
		\caption{A systematic approach to critical review of a corpus of literature}
		\label{fig:graphicalAbstract}
	\end{figure*}
	
	\subsection{Method Summary} \label{subsec:MethodSummary}
	Using thematic analysis with structural coding, a literature review can process a diverse corpus of RUs to produce a succinct, graphical, numerical and analytical representation of the research space. Proceedings may be divided into the following separate tasks. Tasks 1 - 5 are illustrated in \Cref{fig:graphicalAbstract}.
	\begin{enumerate}
		\item Papers are mined (\textbf{\textit{parsed}}) for triads of problems, approaches and developments, using the review protocol described in \cref{subsec:HowAreCodesFormed}.
		\item The codes are clustered around their structural codes, while keeping the links between the codes that organize them into triads.
		\item The clusters are divided into a number of tightly-packed clusters. These clusters are identifiable as categories. 
		\item A category-node is obtained from of each such category, as described in \cref{Categorization}. 
		\item The category-nodes are linked sequentially (horizontally) according to the mined triads to form a causal chain that proceeds from problem to development. The complete set of triads forms the research space’s directed acyclic graph of causality (causality DAG).
		\item Quantitative analysis is possible through the suggested metrics, which provide statistical information on the field. These are complemented by the causality DAG, triads graphic, P-A dyads graphics and taxonomies, which combine to give a multi-faceted profile of the field.
		\item Qualitative analysis is facilitated through the means for grounded reflection provided by the causality DAG and the quantitative analysis.		
	\end{enumerate}
	
	
	
	\bibliography{PAD}
\end{document}